\documentclass[8.5pt,twoside,twocolumn]{article}
\usepackage[super,sort&compress,comma]{natbib} 
\usepackage{mhchem}
\usepackage{times,mathptmx}
\usepackage{sectsty}
\usepackage{balance}

\usepackage{graphicx} 
\usepackage{lastpage}
\usepackage[format=plain,justification=raggedright,singlelinecheck=false,font=small,labelfont=bf,labelsep=space]{caption} 
\usepackage{fancyhdr}
\begin{document}

\thispagestyle{plain}
\fancypagestyle{plain}{
\fancyhead[L]{\includegraphics[height=8pt]{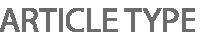}}
\fancyhead[C]{\hspace{-1cm}\includegraphics[height=20pt]{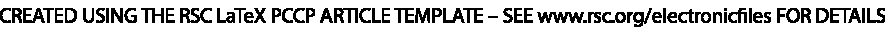}}
\fancyhead[R]{\includegraphics[height=10pt]{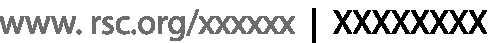}\vspace{-0.2cm}}
\renewcommand{\headrulewidth}{1pt}}
\renewcommand{\thefootnote}{\fnsymbol{footnote}}
\renewcommand\footnoterule{\vspace*{1pt}%
\hrule width 3.4in height 0.4pt \vspace*{5pt}} 
\setcounter{secnumdepth}{5}

\makeatletter 
\def\subsubsection{\@startsection{subsubsection}{3}{10pt}{-1.25ex plus -1ex minus -.1ex}{0ex plus 0ex}{\normalsize\bf}} 
\def\paragraph{\@startsection{paragraph}{4}{10pt}{-1.25ex plus -1ex minus -.1ex}{0ex plus 0ex}{\normalsize\textit}} 
\renewcommand\@biblabel[1]{#1}            
\renewcommand\@makefntext[1]%
{\noindent\makebox[0pt][r]{\@thefnmark\,}#1}
\makeatother 
\renewcommand{\figurename}{\small{Fig.}~}
\sectionfont{\large}
\subsectionfont{\normalsize} 

\fancyfoot{}
\fancyfoot[LO,RE]{\vspace{-7pt}\includegraphics[height=9pt]{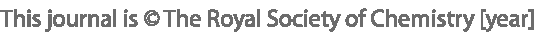}}
\fancyfoot[CO]{\vspace{-7.2pt}\hspace{12.2cm}\includegraphics{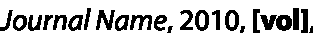}}
\fancyfoot[CE]{\vspace{-7.5pt}\hspace{-13.5cm}\includegraphics{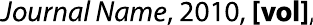}}
\fancyfoot[RO]{\footnotesize{\sffamily{1--\pageref{LastPage} ~\textbar  \hspace{2pt}\thepage}}}
\fancyfoot[LE]{\footnotesize{\sffamily{\thepage~\textbar\hspace{3.45cm} 1--\pageref{LastPage}}}}
\fancyhead{}
\renewcommand{\headrulewidth}{1pt} 
\renewcommand{\footrulewidth}{1pt}
\setlength{\arrayrulewidth}{1pt}
\setlength{\columnsep}{6.5mm}
\setlength\bibsep{1pt}

\twocolumn[
  \begin{@twocolumnfalse}
    \noindent\LARGE{\textbf{Commensurate States and Pattern Switching via Liquid Crystal Skyrmions Trapped in a Square Lattice}}
    \vspace{0.6cm}

\noindent\large{\textbf{A. Duzgun\textit{$^{a}$}, C. Nisoli\textit{$^{a}$}, C. J. O. Reichhardt$^{\ast}$\textit{$^{a}$}, and  C. Reichhardt\textit{$^{a}$}
}}\vspace{0.5cm}

  \noindent\textit{\small{\textbf{Received Xth XXXXXXXXXX 20XX, Accepted Xth XXXXXXXXX 20XX\newline
First published on the web Xth XXXXXXXXXX 200X}}}

\noindent \textbf{\small{DOI: 10.1039/b000000x}}
\vspace{0.6cm}

\noindent\normalsize{
Using continuum based simulations we show that a rich variety of skyrmion liquid crystal states can be realized in the presence of a periodic obstacle array.  As a function of the number of skyrmions per obstacle we find hexagonal, square, dimer, trimer and quadrimer ordering, where the $n$-mer structures are a realization of a molecular crystal state of skyrmions.  As a function of external field and obstacle radius we show that there are transitions between the different crystalline states as well as mixed and disordered structures.  We discuss how these states are related to commensurate effects seen in other systems, such  as vortices in type-II superconductors and colloids interacting with two dimensional substrates.  
}
\vspace{0.5cm}
  \end{@twocolumnfalse}
]

\section{Introduction}

\footnotetext{\textit{$^a$~Theoretical Division and Center for Nonlinear Studies,
    Los Alamos National Laboratory, Los Alamos, New Mexico 87545, USA.
    Fax: 1 505 606 0917; Tel: 1 505 665 1134; E-mail: cjrx@lanl.gov}}

Numerous hard and soft matter systems  can be 
effectively modeled as an assembly of interacting particles coupled to a
two dimensional (2D) periodic substrate.  These  include atoms
and molecules on surfaces~\cite{Coppersmith82,Bak82}, 
vortices in type-II superconductors~\cite{Harada96,Reichhardt00c}
or Bose-Eisenstein condensates~\cite{Pu05,Tung06} 
interacting with periodic pinning arrays,
and charged~\cite{Reichhardt02a,Brunner02,Agra04,Sarlah05,Brazda18} or
magnetic colloids~\cite{Tierno16,Libal18a} on optical traps or structured 
surfaces.  Such systems
exhibit a variety of commensuration effects
in the form of crystalline or superlattice states
when the number of particles 
is an integer multiple of the number of substrate minima.
One example is
the colloidal molecular crystal states found in colloids on 2D arrays,
where  the colloids form localized clusters with synchronized
orientational degrees of freedom~\cite{Reichhardt02a,Brunner02,Agra04,Sarlah05}.
In some cases
plastic crystals~\cite{Reichhardt02a,Brunner02}  can form in which
the number of particles per trap is fixed but there is no
orientational ordering of the clusters.   

Other particle-like objects
are skyrmions,
which arise when the collective behavior of underlying microscopic degrees of freedom
leads to the formation of larger scale structures.
For chiral  magnetic systems, the underlying degrees of freedom
are the spins~\cite{Muhlbauer09,Yu10,Nagaosa13},
while for chiral liquid crystal systems, they
are the molecular director
orientations~\cite{Bogdanov03,Ackerman14,Leonov14,Duzgun18}. 
There has been growing interest in skyrmions and merons in liquid crystals
due to the identification of
new methods to create and control such
systems~\cite{Ackerman14,Fukuda11,Cattaneo16,Nych17,Foster19}.
There is also work examining how liquid crystal (LC)
skyrmions can be manipulated externally~\cite{Sohn19}, 
made to interact with barriers or pinning sites~\cite{Sohn19a,Sohn19b},
or caused to form isolated or collective moving states~\cite{Sohn19a,Ackerman17}. 
    
Since LC skyrmions can form lattices and 
interact with repulsive barriers, it is interesting to examine what types of LC skyrmion
states could be realized when the LC is coupled
a periodic substrate, and compare this  behavior
to the types of ordering found in other systems of particles
coupled to ordered substrates. 
Duzgun and Nisoli~\cite{Duzgun19} recently proposed 
that LC skyrmions interacting repulsively within a square or
triangular array of obstacles can exhibit frustration effects similar to
those found in artificial spin ice systems~\cite{Nisoli13,OrtizAmbriz19}. 
In this work we extend these ideas to examine the types of non-frustrated
commensurate LC skyrmion lattices that can arise when
the skyrmions are coupled to a square lattice of obstacles of the type that 
could  be created by patterned external fields or surface
anchoring~\cite{Sohn19a,Sohn19b,Duzgun19}.  
We find that a rich variety of crystalline states can be stabilized, including
a square lattice at a one to one matching, a dimer
lattice for two skyrmions per obstacle,
and staggered trimer and quadrimer orderings at three and four to one matching.
These $n$-mer states are similar to the colloidal molecular crystal states
observed on periodic substrates~\cite{Reichhardt02a,Brunner02,Agra04,Sarlah05};
however,
the skyrmions exhibit shape distortions that do not occur in the colloidal system.

We show that different structures can be accessed for fixed skyrmion number
when the size of the obstacles is changed
or the external field is varied.
For the case of one to one matching,
we observe a transition from a square to triangular lattice along 
with intermediate states in which the skyrmions form a mixed
structure due to their ability to adopt different sizes
in a single sample.
There can also be pattern switching between different states
when the size of all or a portion of the radii of the obstacles is varied or when
an external field is changed,
suggesting that LC skyrmions on patterned substrates
could exhibit rapid large scale structural transitions that
could be useful for applications.

\section{Numerical Methods}
We model an assembly of $N$ LC skyrmions interacting with a square array 
of $N_{obs}$ obstacles using continuum based simulations~\cite{Duzgun18,Duzgun19}.
We consider 
the traceless tensor $Q$ related to the
scalar order parameter $S$ that quantifies the
orientational order of the  chiral nematic liquid crystal state,
which
gives rise to solutions that  support skyrmions
when the chiral liquid crystal host is confined between two substrates with normal surface anchoring.
The  
free energy
density
has the form 
\begin{align}
f &= (a/2)Tr (Q^2) + (b/3)Tr (Q^3) + (c/4)[(Tr (Q^2)]^2 \nonumber \\
&+ (L/2)(\partial_{\gamma}Q_{\alpha \beta})(\partial_{\gamma}Q_{\alpha \beta}) 
- (4\pi/p)L\epsilon_{\alpha\beta\gamma}Q_{\alpha\rho}{\partial_{\gamma} Q}_{\beta\rho} \nonumber \\
&-[K(\delta(z) + \delta(z -N_{z})) + E^2\Delta \epsilon]Q_{zz}  \nonumber \\
\label{f}
\end{align}
Here the first three terms control the nematic to isotropic transition, the next two terms describe the elastic energies
with respect to a gradient in $Q$,
favoring  a twist with cholesteric pitch $p$,
and the last term is due to the homeotropic surface anchoring at the boundaries
where $K$ is the coupling strength.
The electric field in the $z$-direction is $E$ and $\Delta \epsilon$
is the dielectric anisotropy    
The states are evolved by simulating the following overdamped equation:
\begin{equation}
\frac{\partial Q(r,t)}{\partial t} = -\Gamma \frac{  \delta F}{\delta Q(r,t)}, 
\end{equation}
where $\Gamma$ is the mobility constant
and $F=\int f(x,y,z)\, dx\, dy\, dz  $.
We
denote the field alignment strength as $\alpha=E^2\Delta \epsilon$.
A $z-$invariant structure can be achieved when vertical alignment of molecules is produced solely by the background electric field ({\it i.e.}, by choosing a very small $K\approx 0$ and appropriate value of $\alpha$).
In this work we consider such a $z-$invariant case and model the system as a 3D director ($Q-$tensor)  configuration lying on a 2D surface.
The obstacles are modeled as repulsive barriers of radius $r$
which are realized by applying
an additional electric field within the barrier region that is
much stronger than the background field.
The same effect can be achieved by means of strong surface anchoring localized within
the barriers, which enters the free energy equation identically.
We use the electric field because
it permits dynamic control of the barrier size, shape, and strength.
We first let the system relax,
swell the skyrmion size, and then bring the
skyrmions to a fixed size, which allows for a dynamical annealing effect. Experimentally 
this would be achieved varying the external field.
For the periodic obstacle array we find that a single swell 
cycle is adequate; however, 
for more complex geometries or
a random array, a repeated swell cycle can be applied. 
We focus on a system size of
$4\times 4$
barriers  for filling ratios
of skyrmions to obstacles of $1:1$ up to $4:1$. For specific 
cases we have also considered larger arrays of up to
$20\times 20$ which show the same ground states~\cite{note32}.   

\begin{figure}
  \centering
\includegraphics[width=0.5\textwidth]{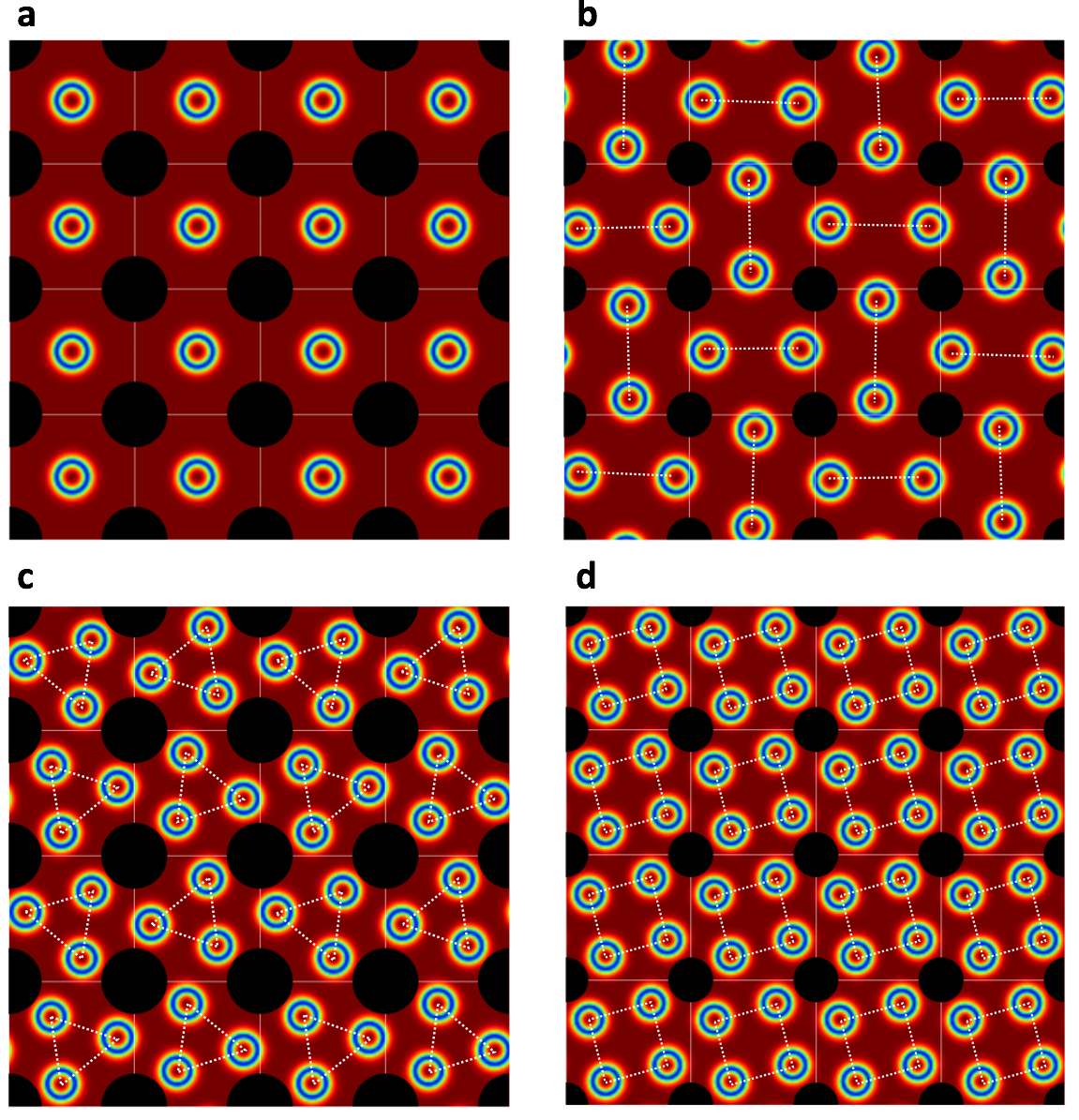}
  \caption{The liquid crystal skyrmions (blue rings) and
    obstacle locations (black circles) obtained from a continuum based simulation of a 
    chiral liquid crystal state.
    In each case $\alpha  = 0.3$.
    The colors represent the orientation of the director field.
    (a) The 1:1 matching at an obstacle radius of $r = 15$ showing a square skyrmion lattice.
    (b) The 2:1 state at $r = 10$ with an alternating dimer ordering as
    indicated by the white dashed lines. 
    (c) The 3:1 state at $r = 15$ where there is trimer ordering with a
    small shift at every other plaquette.
    (d) The 4:1 state showing a quadrimer ordering at $r = 10$.
    Here the size of the square cells is $L=60$.
}
\label{fig:1}
\end{figure}

\section{Results}
In Fig.~\ref{fig:1}(a) we show the location of the barriers of spacing $L=60$
and the skyrmions
at $r = 15$.
Here there is a 1:1 matching of
the number of skyrmions to the number of obstacles.
The skyrmions form a square lattice  located
in the center of the interstitial regions between the obstacles.
Figure~\ref{fig:1}(b) shows the
case of two skyrmions per obstacle with $r= 10$, where the system
forms a dimer lattice as indicated by the dashed lines connecting pairs
of skyrmions.  The dimers have an additional long range orientational ordering,
and each dimer is alternately vertical or horizontal.
In Fig.~\ref{fig:1}(c),  at the $3:1$ matching with $r = 15$,
the skyrmions form an ordered trimer state as indicated by the lines.
The trimers exhibit an additional
small canting from one plaquette to the next. 
At $4:1$ with $r =10$ in Fig.~\ref{fig:1}(d),  the skyrmions form a quadrimer state which  
is the same for each plaquette.
These states are similar to the $N$-mer orderings found in colloidal molecular crystal
systems for colloids interacting with square or triangular
substrates~\cite{Reichhardt02a,Brunner02,Agra04,Sarlah05}.
In particular, the
dimer state has been
described as an anti-ferromagnetic Ising model on a square lattice~\cite{Agra04}, where the
orientation of the dimer corresponds to the two possible
orientations of an effective spin.
A similar dimer state was also predicted for vortices in a Bose-Einstein condensate
on a square
lattice at the second matching filling~\cite{Pu05}.
The trimer ordering in the colloidal system~\cite{Reichhardt02a} differs from that
for the skyrmions in Fig.~\ref{fig:1}(c).
The colloidal trimers have stripe or columnar
orientational ordering
due to the longer range
multipolar charge interaction between trimers,
whereas the LC skyrmion trimers experience only short range repulsion
and have only weak orientational ordering along the horizontal direction.
The LC skyrmion state has the same ordering as the colloidal state at the
fourth filling for the square lattice~\cite{Reichhardt02a}. 
We call the states 
in Fig.~\ref{fig:1}(b,c,d) LC skyrmion molecular crystals since the $N$-mers have both positional
and orientational order.

\begin{figure}
  \centering
\includegraphics[width=0.5\textwidth]{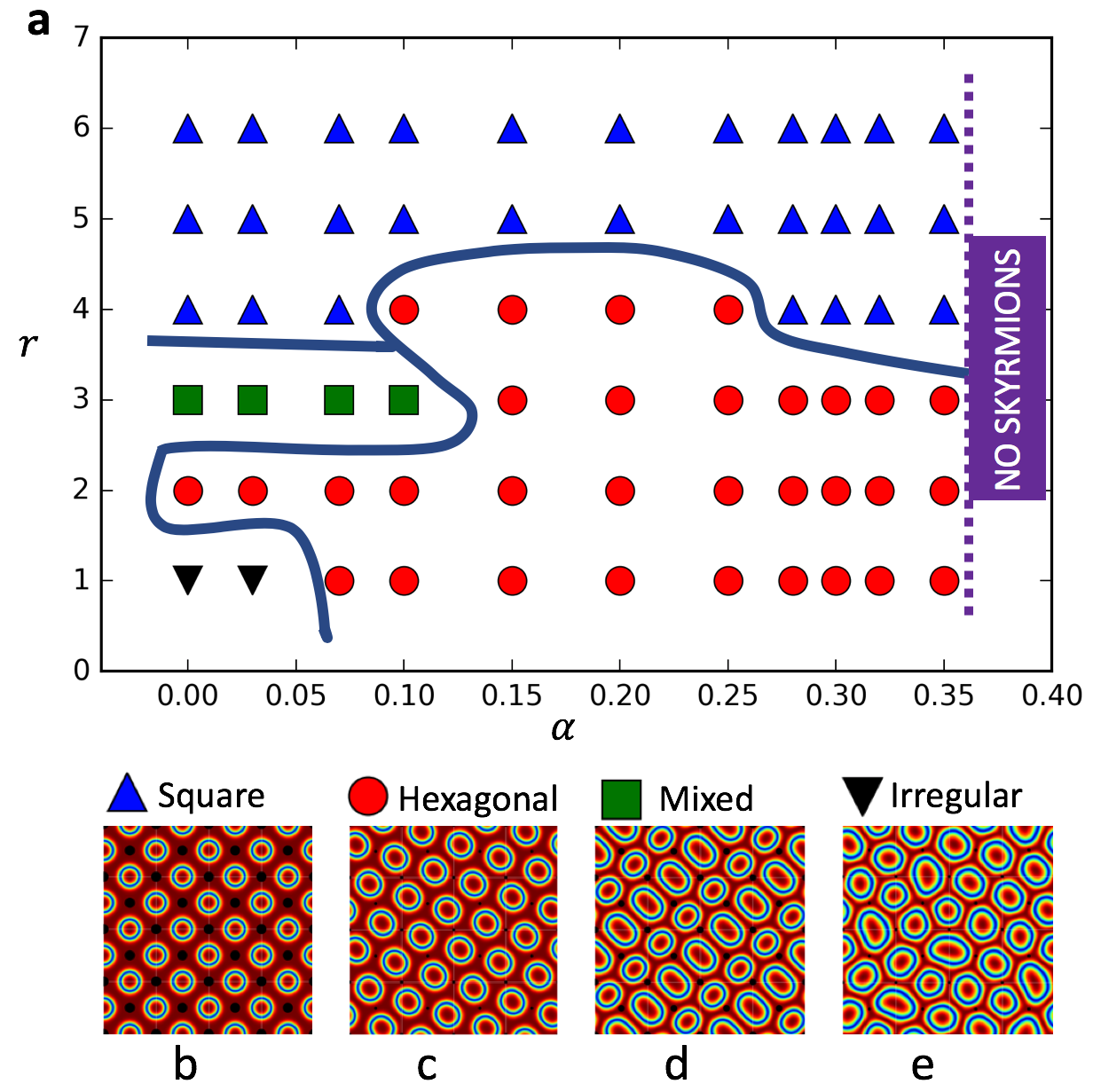}
\caption{
  (a) The phase diagram as a function of obstacle size $r$
    vs the background field $\alpha$ for the system in Fig.~\ref{fig:1}(a) at a 1:1 matching
    of LC skyrmions to obstacles.
    Blue triangles:  square lattice states, shown in panel (b);
    red squares: hexagonal lattices, shown in panel (c);
    green squares: mixed state, shown in panel (d);
    black triangles:  disordered or irregular states, shown in panel (e).
    For fields greater than $\alpha=0.35$, skyrmions
do not form, while for $6.0 < r < 16.0$ there is only a square lattice state.
}
\label{fig:2}
\end{figure}

We next consider the effect of changing the background field and the obstacle
radius for the $1:1$ filling.
For vortices and other particle based systems on a square array at the first matching
filling, 
there can be a transition
from a square lattice at strong coupling where the substrate dominates
the behavior to a triangular lattice at weak coupling where the
particle-particle interaction dominates
the behavior~\cite{Reichhardt01c,Granz16}.
For hard disks at a  $1:1$ matching on a square lattice, there can be a transition
to a hexagonal lattice and even a rhombic phase
as a function of substrate strength and disk size~\cite{Neuhaus13}.
In Fig.~\ref{fig:2}(a) we show the phase diagram as a function of
the obstacle radius $r$ versus background field $\alpha$ for the $1:1$
filling, where
we observe five phases.
For large fields $\alpha>0.35$,
skyrmions do not appear and the system has a uniform background. 
For large defect radius $r > 5.0$, there is an extended region in which
the system forms
a commensurate square lattice, as
illustrated in Fig.~\ref{fig:2}(b).
The square lattice extends over the range
$6.0 < r < 16.0$, 
but we focus on the regime $r < 7.0$ since
this is where additional phases occur.
When $r$ is small but $\alpha$ is large, 
the skyrmions have more room to distort,
allowing them to form a hexagonal lattice as shown in
Fig.~\ref{fig:2}(c).
There is a window at $r=3$ where, for
small $\alpha$,
the skyrmions can more easily change shape to create a mixed state as illustrated
in Fig.~\ref{fig:2}(d).
In this case, half of the skyrmions
become elongated and the overall pattern has a superlattice ordering.
At small obstacle radius and
small field we find a disordered state with skyrmions
in a mixture of sizes, as shown in Fig.~\ref{fig:2}(e).
For other fillings, a similar phase diagram can be constructed,
where obstacles of large size produce
$N$-mer states,
while
disordered or hexagonal lattices generally form
for smaller $r$.

\begin{figure}
  \centering
\includegraphics[width=0.5\textwidth]{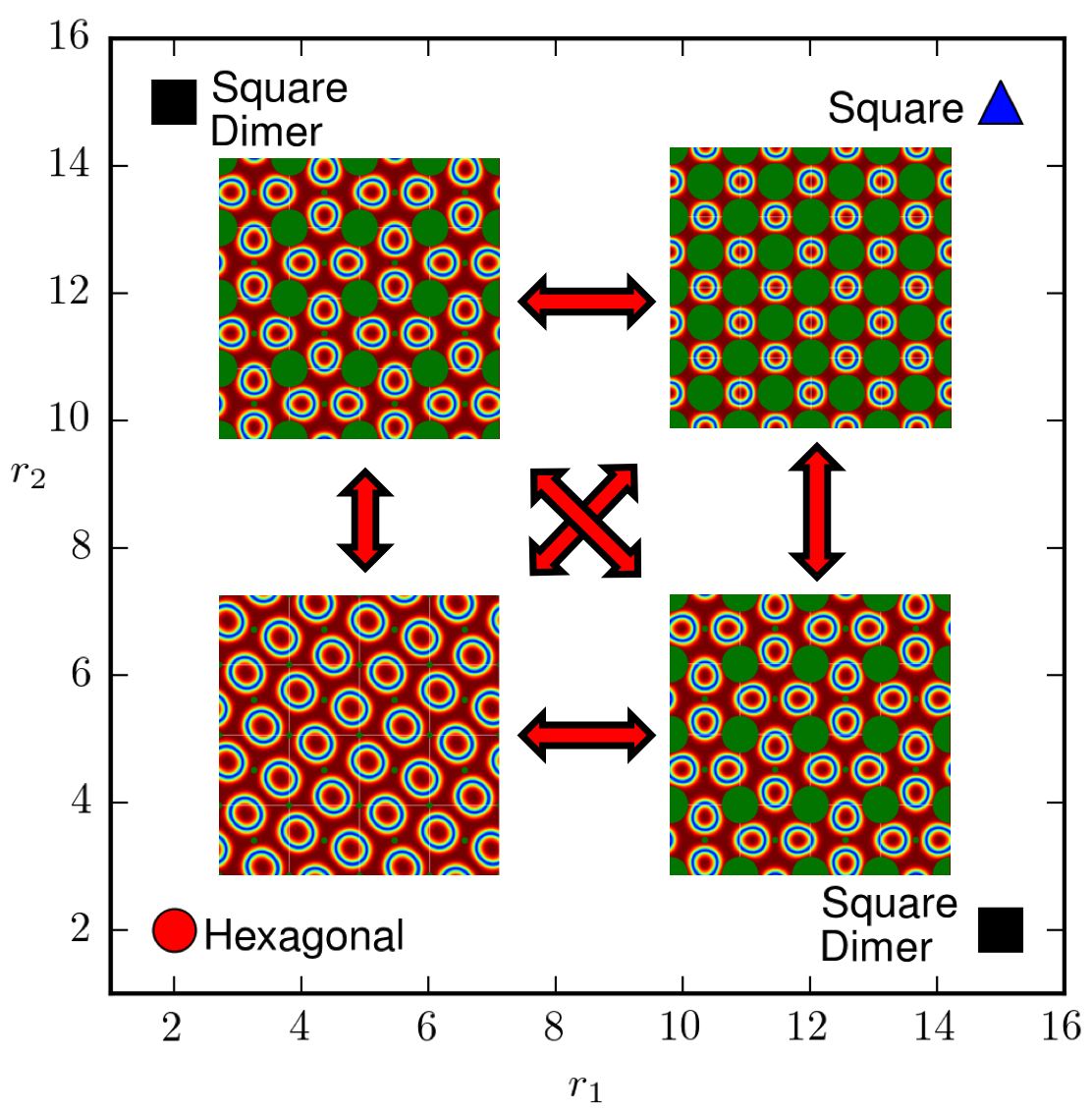}
  \caption{ The different possible states for LC skyrmions
    on a square lattice at a 1:1 filling where the obstacles have
    two different radii, $r_{1}$ and $r_{2}$.
    When $r_{1}$ and $r_{2}$ are both large, the system forms a square lattice
    (upper left).
    When $r_{1}$ and $r_{2}$ are both
    small,
    hexagonal ordering emerges (lower right).
    For $r_1\gg r_2$ or $r_2 \gg r_1$ (lower left and upper right),
    a dimer lattice emerges.
    The arrows indicate the different routes that could be taken to get from one
state to another.  
}
\label{fig:3}
\end{figure}

The fact that different patterns can arise as a function of obstacle size
suggests that various types of pattern
switching could be achieved by suddenly changing the sizes of all or
a portion of the obstacles.
An example of how this could be achieved for fixed skyrmion number is shown in
Fig.~\ref{fig:3}, 
where half of the obstacles have radius 
$r_{1}$ and the other half have radius $r_{2}$.
Here the background field is fixed at $\alpha=0.2$. 
When $r_{1}$ and $r_{2}$ are both large, the system forms a square lattice
as shown in the upper right panel.
If $r_{1}$ and $r_{2}$ are both small,
a hexagonal lattice appears as illustrated
in the lower left panel, while for the cases 
of $r_{1}$ much smaller than $r_{2}$ or $r_{2}$ much smaller than $r_{1}$,
the system forms the 
dimer lattice indicated in the upper left and lower right corners.
The arrows indicate the possible routes along which the
different patterns could be switched. 
This suggests that LC skyrmions interacting with ordered structures
can undergo large scale switching behaviors that could be useful
for creating devices. 

\begin{figure}
  \centering
\includegraphics[width=0.5\textwidth]{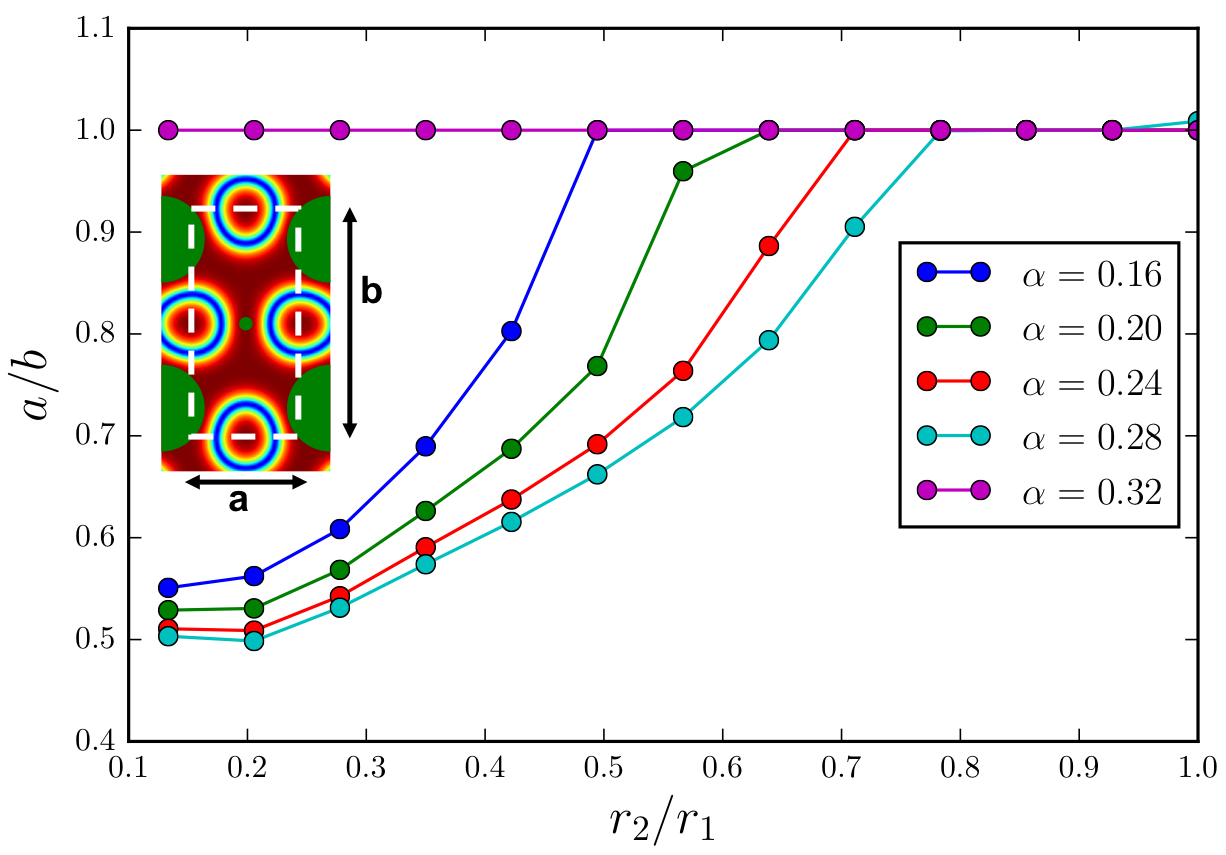}
\caption{
  The ratio of the sides $a/b$ (shown in the inset) vs $r_{2}/r_{1}$
  for samples from Fig.~\ref{fig:3} with $1:1$ matching 
  at $r_{1} = 15$.
  When $a/b = 1.0$, the system forms a square lattice,
  while for $a/b \approx 0.5$, an ordered dimer lattice appears.
  For varied $\alpha$ or changing obstacle size,
  transitions occur between the two states.  
}
\label{fig:4}
\end{figure}

In Fig.~\ref{fig:4} we plot the ratio of the distance between skyrmions,
as indicated in the rectangular box in the inset, of 
side $a$ and side $b$ versus $r_{2}/r_{1}$ for the
system in Fig.~\ref{fig:3}, where we fix $r_1 = 15$ and vary $r_{2}$.
We show results for external field
values of $\alpha= 0.16$ to $\alpha=0.32$.
When $a/b = 1.0$, we find a square lattice, while for
$a/b = 0.5$, a dimer lattice appears.
Upon 
increasing $\alpha$, the transition to the square lattice shifts to
higher 
ratios of $r_2/r_1$; however, at $\alpha = 0.32$ the system remains
in the square lattice for all values of $r_{1}/r_{2}$ since the
skyrmions are so small that they no longer interact with one another
and show little distortion from their initial positions.
This indicates that not only can the skyrmion pattern be switched,
but also the geometric ratio of the pattern
can be controlled as function of the electric field.  

\section{Discussion}
Our results should be general for even higher fillings $N$, leading to higher order 
$N$-mer states. For finite thermal fluctuations,
the different states could show additional effects such as a transition
from  a molecular crystal
state to a plastic crystal state in which the $N$-mers are randomly rotating,
similar to what has been  observed in colloidal molecular crystal
systems~\cite{Reichhardt02a,Brunner02,Agra04,Sarlah05}. 
Beyond commensurate states,
there should also be a  number of incommensurate states at non-integer matchings,
which could form partially disordered
states or rational commensurate states
when the ratio of skyrmions to obstacles is rational.
In particle based systems, incommensurate states form 
kinks or antikinks~\cite{Bohlein12,Vanossi12,McDermott13}.
In the LC skyrmion system, however,
due to the ability of the skyrmions to change size, the 
kinks can shrink or expand 
in order to reduce the energy cost of the defect,
so that LC skyrmions could be 
much more robust to disordering due to incommensurations.
These results could also be extended to other obstacle lattice geometries 
such as triangular lattices, mixed lattices, quasiperiodic lattices,
or random arrays.

In commensurate-incommensurate systems,
a variety of dynamics can arise~\cite{Bohlein12,Vanossi12} when the system is driven.
Driving of LC systems has already been demonstrated~\cite{Ackerman17},
so the dynamics of the skyrmions could be explored for driving over a
periodic substrate.
Finally, similar states could arise for magnetic skyrmions
coupled to a periodic obstacle array, and there are already proposals
on how to create
such substrates for magnetic skyrmions on an anti-dot lattice~\cite{Feilhauer19}.

\section{Summary}
We have used continuum based simulations to examine the ordering of liquid crystal
skyrmions interacting with a square obstacle array which could be created
using anchoring or with fields.
As a function of filling, we find that a variety of
crystalline states can be stabilized,
including a square lattice, an alternating dimer lattice,
a trimer state, and a quadrimer state.
We refer to
the dimer and higher order $N$-mer states as skyrmion LC molecular crystal states
in analogy to
colloidal molecular crystals.
For the commensurate $1:1$ filling, we map out the phase diagram as 
a function of barrier size and field
and show that five different phases arise:
no skyrmions, a square lattice, a hexagonal lattice, a disordered state,
and a mixed phase.
The mixed phase consists of a superlattice of skyrmions of different sizes.
We also show that the system can exhibit pattern switching 
between dimer, hexagonal and square lattices as a function
of the ratio of the obstacle size to the external field.
We discuss future directions such as
incommensurate states, other obstacle lattice geometries, and driving.
Liquid crystal skyrmions
represent another system that can be used to realize commensurate states 
for an assembly of particle-like objects
coupled to a periodic substrate.

\section{Acknowledgements}
We gratefully acknowledge the support of the U.S. Department of
Energy through the LANL/LDRD program for this work.
This work was carried out under the auspices of the 
NNSA of the 
U.S. DoE
at 
LANL
under Contract No.
DE-AC52-06NA25396 and through the LANL/LDRD program.

\footnotesize{
  \bibliography{mybib}

\providecommand*{\mcitethebibliography}{\thebibliography}
\csname @ifundefined\endcsname{endmcitethebibliography}
{\let\endmcitethebibliography\endthebibliography}{}
\begin{mcitethebibliography}{39}
\providecommand*{\natexlab}[1]{#1}
\providecommand*{\mciteSetBstSublistMode}[1]{}
\providecommand*{\mciteSetBstMaxWidthForm}[2]{}
\providecommand*{\mciteBstWouldAddEndPuncttrue}
  {\def\EndOfBibitem{\unskip.}}
\providecommand*{\mciteBstWouldAddEndPunctfalse}
  {\let\EndOfBibitem\relax}
\providecommand*{\mciteSetBstMidEndSepPunct}[3]{}
\providecommand*{\mciteSetBstSublistLabelBeginEnd}[3]{}
\providecommand*{\EndOfBibitem}{}
\mciteSetBstSublistMode{f}
\mciteSetBstMaxWidthForm{subitem}
{(\emph{\alph{mcitesubitemcount}})}
\mciteSetBstSublistLabelBeginEnd{\mcitemaxwidthsubitemform\space}
{\relax}{\relax}

\bibitem[Coppersmith \emph{et~al.}(1982)Coppersmith, Fisher, Halperin, Lee, and
  Brinkman]{Coppersmith82}
S.~N. Coppersmith, D.~S. Fisher, B.~I. Halperin, P.~A. Lee and W.~F. Brinkman,
  \emph{Phys. Rev. B}, 1982, \textbf{25}, 349--363\relax
\mciteBstWouldAddEndPuncttrue
\mciteSetBstMidEndSepPunct{\mcitedefaultmidpunct}
{\mcitedefaultendpunct}{\mcitedefaultseppunct}\relax
\EndOfBibitem
\bibitem[Bak(1982)]{Bak82}
P.~Bak, \emph{Rep. Prog. Phys.}, 1982, \textbf{45}, 587--629\relax
\mciteBstWouldAddEndPuncttrue
\mciteSetBstMidEndSepPunct{\mcitedefaultmidpunct}
{\mcitedefaultendpunct}{\mcitedefaultseppunct}\relax
\EndOfBibitem
\bibitem[Harada \emph{et~al.}(1996)Harada, Kamimura, Kasai, Matsuda, Tonomura,
  and Moshchalkov]{Harada96}
K.~Harada, O.~Kamimura, H.~Kasai, T.~Matsuda, A.~Tonomura and V.~V.
  Moshchalkov, \emph{Science}, 1996, \textbf{274}, 1167--1170\relax
\mciteBstWouldAddEndPuncttrue
\mciteSetBstMidEndSepPunct{\mcitedefaultmidpunct}
{\mcitedefaultendpunct}{\mcitedefaultseppunct}\relax
\EndOfBibitem
\bibitem[Reichhardt and Gr\o{}nbech-Jensen(2000)]{Reichhardt00c}
C.~Reichhardt and N.~Gr\o{}nbech-Jensen, \emph{Phys. Rev. Lett.}, 2000,
  \textbf{85}, 2372--2375\relax
\mciteBstWouldAddEndPuncttrue
\mciteSetBstMidEndSepPunct{\mcitedefaultmidpunct}
{\mcitedefaultendpunct}{\mcitedefaultseppunct}\relax
\EndOfBibitem
\bibitem[Pu \emph{et~al.}(2005)Pu, Baksmaty, Yi, and Bigelow]{Pu05}
H.~Pu, L.~O. Baksmaty, S.~Yi and N.~P. Bigelow, \emph{Phys. Rev. Lett.}, 2005,
  \textbf{94}, 190401\relax
\mciteBstWouldAddEndPuncttrue
\mciteSetBstMidEndSepPunct{\mcitedefaultmidpunct}
{\mcitedefaultendpunct}{\mcitedefaultseppunct}\relax
\EndOfBibitem
\bibitem[Tung \emph{et~al.}(2006)Tung, Schweikhard, and Cornell]{Tung06}
S.~Tung, V.~Schweikhard and E.~A. Cornell, \emph{Phys. Rev. Lett.}, 2006,
  \textbf{97}, 240402\relax
\mciteBstWouldAddEndPuncttrue
\mciteSetBstMidEndSepPunct{\mcitedefaultmidpunct}
{\mcitedefaultendpunct}{\mcitedefaultseppunct}\relax
\EndOfBibitem
\bibitem[Reichhardt and Olson(2002)]{Reichhardt02a}
C.~Reichhardt and C.~J. Olson, \emph{Phys. Rev. Lett.}, 2002, \textbf{88},
  248301\relax
\mciteBstWouldAddEndPuncttrue
\mciteSetBstMidEndSepPunct{\mcitedefaultmidpunct}
{\mcitedefaultendpunct}{\mcitedefaultseppunct}\relax
\EndOfBibitem
\bibitem[Brunner and Bechinger(2002)]{Brunner02}
M.~Brunner and C.~Bechinger, \emph{Phys. Rev. Lett.}, 2002, \textbf{88},
  248302\relax
\mciteBstWouldAddEndPuncttrue
\mciteSetBstMidEndSepPunct{\mcitedefaultmidpunct}
{\mcitedefaultendpunct}{\mcitedefaultseppunct}\relax
\EndOfBibitem
\bibitem[Agra \emph{et~al.}(2004)Agra, van Wijland, and Trizac]{Agra04}
R.~Agra, F.~van Wijland and E.~Trizac, \emph{Phys. Rev. Lett.}, 2004,
  \textbf{93}, 018304\relax
\mciteBstWouldAddEndPuncttrue
\mciteSetBstMidEndSepPunct{\mcitedefaultmidpunct}
{\mcitedefaultendpunct}{\mcitedefaultseppunct}\relax
\EndOfBibitem
\bibitem[\ifmmode~\check{S}\else \v{S}\fi{}arlah
  \emph{et~al.}(2005)\ifmmode~\check{S}\else \v{S}\fi{}arlah, Franosch, and
  Frey]{Sarlah05}
A.~\ifmmode~\check{S}\else \v{S}\fi{}arlah, T.~Franosch and E.~Frey,
  \emph{Phys. Rev. Lett.}, 2005, \textbf{95}, 088302\relax
\mciteBstWouldAddEndPuncttrue
\mciteSetBstMidEndSepPunct{\mcitedefaultmidpunct}
{\mcitedefaultendpunct}{\mcitedefaultseppunct}\relax
\EndOfBibitem
\bibitem[Brazda \emph{et~al.}(2018)Brazda, Silva, Manini, Vanossi, Guerra,
  Tosatti, and Bechinger]{Brazda18}
T.~Brazda, A.~Silva, N.~Manini, A.~Vanossi, R.~Guerra, E.~Tosatti and
  C.~Bechinger, \emph{Phys. Rev. X}, 2018, \textbf{8}, 011050\relax
\mciteBstWouldAddEndPuncttrue
\mciteSetBstMidEndSepPunct{\mcitedefaultmidpunct}
{\mcitedefaultendpunct}{\mcitedefaultseppunct}\relax
\EndOfBibitem
\bibitem[Tierno(2016)]{Tierno16}
P.~Tierno, \emph{Phys. Rev. Lett.}, 2016, \textbf{116}, 038303\relax
\mciteBstWouldAddEndPuncttrue
\mciteSetBstMidEndSepPunct{\mcitedefaultmidpunct}
{\mcitedefaultendpunct}{\mcitedefaultseppunct}\relax
\EndOfBibitem
\bibitem[Lib{\' a}l \emph{et~al.}(2018)Lib{\' a}l, Lee, Ortiz-Ambriz,
  Reichhardt, Reichhardt, Tierno, and Nisoli]{Libal18a}
A.~Lib{\' a}l, D.~Y. Lee, A.~Ortiz-Ambriz, C.~Reichhardt, C.~J.~O. Reichhardt,
  P.~Tierno and C.~Nisoli, \emph{Nature Commun.}, 2018, \textbf{9}, 4146\relax
\mciteBstWouldAddEndPuncttrue
\mciteSetBstMidEndSepPunct{\mcitedefaultmidpunct}
{\mcitedefaultendpunct}{\mcitedefaultseppunct}\relax
\EndOfBibitem
\bibitem[M{\" u}hlbauer \emph{et~al.}(2009)M{\" u}hlbauer, Binz, Jonietz,
  Pfleiderer, Rosch, Neubauer, Georgii, and B{\" o}ni]{Muhlbauer09}
S.~M{\" u}hlbauer, B.~Binz, F.~Jonietz, C.~Pfleiderer, A.~Rosch, A.~Neubauer,
  R.~Georgii and P.~B{\" o}ni, \emph{Science}, 2009, \textbf{323},
  915--919\relax
\mciteBstWouldAddEndPuncttrue
\mciteSetBstMidEndSepPunct{\mcitedefaultmidpunct}
{\mcitedefaultendpunct}{\mcitedefaultseppunct}\relax
\EndOfBibitem
\bibitem[Yu \emph{et~al.}(2010)Yu, Onose, Kanazawa, Park, Han, Matsui, Nagaosa,
  and Tokura]{Yu10}
X.~Z. Yu, Y.~Onose, N.~Kanazawa, J.~H. Park, J.~H. Han, Y.~Matsui, N.~Nagaosa
  and Y.~Tokura, \emph{Nature (London)}, 2010, \textbf{465}, 901--904\relax
\mciteBstWouldAddEndPuncttrue
\mciteSetBstMidEndSepPunct{\mcitedefaultmidpunct}
{\mcitedefaultendpunct}{\mcitedefaultseppunct}\relax
\EndOfBibitem
\bibitem[Nagaosa and Tokura(2013)]{Nagaosa13}
N.~Nagaosa and Y.~Tokura, \emph{Nature Nanotechnol.}, 2013, \textbf{8},
  899--911\relax
\mciteBstWouldAddEndPuncttrue
\mciteSetBstMidEndSepPunct{\mcitedefaultmidpunct}
{\mcitedefaultendpunct}{\mcitedefaultseppunct}\relax
\EndOfBibitem
\bibitem[Bogdanov \emph{et~al.}(2003)Bogdanov, R\"o\ss{}ler, and
  Shestakov]{Bogdanov03}
A.~N. Bogdanov, U.~K. R\"o\ss{}ler and A.~A. Shestakov, \emph{Phys. Rev. E},
  2003, \textbf{67}, 016602\relax
\mciteBstWouldAddEndPuncttrue
\mciteSetBstMidEndSepPunct{\mcitedefaultmidpunct}
{\mcitedefaultendpunct}{\mcitedefaultseppunct}\relax
\EndOfBibitem
\bibitem[Ackerman \emph{et~al.}(2014)Ackerman, Trivedi, Senyuk, van~de
  Lagemaat, and Smalyukh]{Ackerman14}
P.~J. Ackerman, R.~P. Trivedi, B.~Senyuk, J.~van~de Lagemaat and I.~I.
  Smalyukh, \emph{Phys. Rev. E}, 2014, \textbf{90}, 012505\relax
\mciteBstWouldAddEndPuncttrue
\mciteSetBstMidEndSepPunct{\mcitedefaultmidpunct}
{\mcitedefaultendpunct}{\mcitedefaultseppunct}\relax
\EndOfBibitem
\bibitem[Leonov \emph{et~al.}(2014)Leonov, Dragunov, R\"o\ss{}ler, and
  Bogdanov]{Leonov14}
A.~O. Leonov, I.~E. Dragunov, U.~K. R\"o\ss{}ler and A.~N. Bogdanov,
  \emph{Phys. Rev. E}, 2014, \textbf{90}, 042502\relax
\mciteBstWouldAddEndPuncttrue
\mciteSetBstMidEndSepPunct{\mcitedefaultmidpunct}
{\mcitedefaultendpunct}{\mcitedefaultseppunct}\relax
\EndOfBibitem
\bibitem[Duzgun \emph{et~al.}(2018)Duzgun, Selinger, and Saxena]{Duzgun18}
A.~Duzgun, J.~V. Selinger and A.~Saxena, \emph{Phys. Rev. E}, 2018,
  \textbf{97}, 062706\relax
\mciteBstWouldAddEndPuncttrue
\mciteSetBstMidEndSepPunct{\mcitedefaultmidpunct}
{\mcitedefaultendpunct}{\mcitedefaultseppunct}\relax
\EndOfBibitem
\bibitem[Fukuda and \v{Z}umer(2011)]{Fukuda11}
J.~Fukuda and S.~\v{Z}umer, \emph{Nature Commun.}, 2011, \textbf{2}, 246\relax
\mciteBstWouldAddEndPuncttrue
\mciteSetBstMidEndSepPunct{\mcitedefaultmidpunct}
{\mcitedefaultendpunct}{\mcitedefaultseppunct}\relax
\EndOfBibitem
\bibitem[Cattaneo \emph{et~al.}(2016)Cattaneo, Kos, Savoini, Kouwer, Rowan,
  Ravnik, Mu\v{s}evi\v{c}, and Rasing]{Cattaneo16}
L.~Cattaneo, v.~Kos, M.~Savoini, P.~Kouwer, A.~Rowan, M.~Ravnik,
  I.~Mu\v{s}evi\v{c} and T.~Rasing, \emph{Soft Matter}, 2016, \textbf{12},
  853--858\relax
\mciteBstWouldAddEndPuncttrue
\mciteSetBstMidEndSepPunct{\mcitedefaultmidpunct}
{\mcitedefaultendpunct}{\mcitedefaultseppunct}\relax
\EndOfBibitem
\bibitem[Nych \emph{et~al.}(2017)Nych, Fukuda, Ognysta, \v{Z}umer, and
  Mu\v{s}evi\v{c}]{Nych17}
A.~Nych, J.~Fukuda, U.~Ognysta, S.~\v{Z}umer and I.~Mu\v{s}evi\v{c},
  \emph{Nature Phys.}, 2017, \textbf{13}, 1215\relax
\mciteBstWouldAddEndPuncttrue
\mciteSetBstMidEndSepPunct{\mcitedefaultmidpunct}
{\mcitedefaultendpunct}{\mcitedefaultseppunct}\relax
\EndOfBibitem
\bibitem[Foster \emph{et~al.}(2019)Foster, Kind, Ackerman, Tai, Dennis, and
  Smalyukh]{Foster19}
D.~Foster, C.~Kind, P.~J. Ackerman, J.-S.~B. Tai, M.~R. Dennis and I.~I.
  Smalyukh, \emph{Nature Phys.}, 2019, \textbf{15}, 655\relax
\mciteBstWouldAddEndPuncttrue
\mciteSetBstMidEndSepPunct{\mcitedefaultmidpunct}
{\mcitedefaultendpunct}{\mcitedefaultseppunct}\relax
\EndOfBibitem
\bibitem[Sohn \emph{et~al.}(2019)Sohn, Liu, Wang, and Smalyukh]{Sohn19}
H.~R.~O. Sohn, C.~D. Liu, Y.~Wang and I.~I. Smalyukh, \emph{Optics Express},
  2019, \textbf{27}, 29055--29068\relax
\mciteBstWouldAddEndPuncttrue
\mciteSetBstMidEndSepPunct{\mcitedefaultmidpunct}
{\mcitedefaultendpunct}{\mcitedefaultseppunct}\relax
\EndOfBibitem
\bibitem[Sohn \emph{et~al.}(2019)Sohn, Liu, and Smalyukh]{Sohn19a}
H.~R.~O. Sohn, C.~D. Liu and I.~I. Smalyukh, \emph{Nature Commun.}, 2019,
  \textbf{10}, 4744\relax
\mciteBstWouldAddEndPuncttrue
\mciteSetBstMidEndSepPunct{\mcitedefaultmidpunct}
{\mcitedefaultendpunct}{\mcitedefaultseppunct}\relax
\EndOfBibitem
\bibitem[Sohn \emph{et~al.}()Sohn, Liu, Voinescu, Chen, and Smalyukh]{Sohn19b}
H.~R.~O. Sohn, C.~D. Liu, R.~Voinescu, Z.~Chen and I.~I. Smalyukh,
  {arXiv:1911.04640}\relax
\mciteBstWouldAddEndPuncttrue
\mciteSetBstMidEndSepPunct{\mcitedefaultmidpunct}
{\mcitedefaultendpunct}{\mcitedefaultseppunct}\relax
\EndOfBibitem
\bibitem[Ackerman \emph{et~al.}(2017)Ackerman, Boyle, and Smalyukh]{Ackerman17}
P.~J. Ackerman, T.~Boyle and I.~I. Smalyukh, \emph{Nature Commun.}, 2017,
  \textbf{8}, 673\relax
\mciteBstWouldAddEndPuncttrue
\mciteSetBstMidEndSepPunct{\mcitedefaultmidpunct}
{\mcitedefaultendpunct}{\mcitedefaultseppunct}\relax
\EndOfBibitem
\bibitem[Duzgun and Nisoli()]{Duzgun19}
A.~Duzgun and C.~Nisoli, {arXiv:1908.03246}\relax
\mciteBstWouldAddEndPuncttrue
\mciteSetBstMidEndSepPunct{\mcitedefaultmidpunct}
{\mcitedefaultendpunct}{\mcitedefaultseppunct}\relax
\EndOfBibitem
\bibitem[Nisoli \emph{et~al.}(2013)Nisoli, Moessner, and Schiffer]{Nisoli13}
C.~Nisoli, R.~Moessner and P.~Schiffer, \emph{Rev. Mod. Phys.}, 2013,
  \textbf{85}, 1473--1490\relax
\mciteBstWouldAddEndPuncttrue
\mciteSetBstMidEndSepPunct{\mcitedefaultmidpunct}
{\mcitedefaultendpunct}{\mcitedefaultseppunct}\relax
\EndOfBibitem
\bibitem[Ortiz-Ambriz \emph{et~al.}()Ortiz-Ambriz, Nisoli, Reichhardt,
  Reichhardt, and Tierno]{OrtizAmbriz19}
A.~Ortiz-Ambriz, C.~Nisoli, C.~Reichhardt, C.~J.~O. Reichhardt and P.~Tierno,
  {arXiv:1909.13534}\relax
\mciteBstWouldAddEndPuncttrue
\mciteSetBstMidEndSepPunct{\mcitedefaultmidpunct}
{\mcitedefaultendpunct}{\mcitedefaultseppunct}\relax
\EndOfBibitem
\bibitem[not()]{note32}
See supplementary information for the larger system sizes.\relax
\mciteBstWouldAddEndPunctfalse
\mciteSetBstMidEndSepPunct{\mcitedefaultmidpunct}
{}{\mcitedefaultseppunct}\relax
\EndOfBibitem
\bibitem[Reichhardt \emph{et~al.}(2001)Reichhardt, Olson, Scalettar, and
  Zim\'anyi]{Reichhardt01c}
C.~Reichhardt, C.~J. Olson, R.~T. Scalettar and G.~T. Zim\'anyi, \emph{Phys.
  Rev. B}, 2001, \textbf{64}, 144509\relax
\mciteBstWouldAddEndPuncttrue
\mciteSetBstMidEndSepPunct{\mcitedefaultmidpunct}
{\mcitedefaultendpunct}{\mcitedefaultseppunct}\relax
\EndOfBibitem
\bibitem[Gr\"anz \emph{et~al.}(2016)Gr\"anz, Korshunov, Geshkenbein, and
  Blatter]{Granz16}
B.~Gr\"anz, S.~E. Korshunov, V.~B. Geshkenbein and G.~Blatter, \emph{Phys. Rev.
  B}, 2016, \textbf{94}, 054110\relax
\mciteBstWouldAddEndPuncttrue
\mciteSetBstMidEndSepPunct{\mcitedefaultmidpunct}
{\mcitedefaultendpunct}{\mcitedefaultseppunct}\relax
\EndOfBibitem
\bibitem[Neuhaus \emph{et~al.}(2013)Neuhaus, Marechal, Schmiedeberg, and
  L\"owen]{Neuhaus13}
T.~Neuhaus, M.~Marechal, M.~Schmiedeberg and H.~L\"owen, \emph{Phys. Rev.
  Lett.}, 2013, \textbf{110}, 118301\relax
\mciteBstWouldAddEndPuncttrue
\mciteSetBstMidEndSepPunct{\mcitedefaultmidpunct}
{\mcitedefaultendpunct}{\mcitedefaultseppunct}\relax
\EndOfBibitem
\bibitem[Bohlein \emph{et~al.}(2012)Bohlein, Mikhael, and Bechinger]{Bohlein12}
T.~Bohlein, J.~Mikhael and C.~Bechinger, \emph{Nature Mater.}, 2012,
  \textbf{11}, 126--130\relax
\mciteBstWouldAddEndPuncttrue
\mciteSetBstMidEndSepPunct{\mcitedefaultmidpunct}
{\mcitedefaultendpunct}{\mcitedefaultseppunct}\relax
\EndOfBibitem
\bibitem[Vanossi \emph{et~al.}(2012)Vanossi, Manini, and Tosatti]{Vanossi12}
A.~Vanossi, N.~Manini and E.~Tosatti, \emph{Proc. Natl. Acad. Sci. (USA)},
  2012, \textbf{109}, 16429--16433\relax
\mciteBstWouldAddEndPuncttrue
\mciteSetBstMidEndSepPunct{\mcitedefaultmidpunct}
{\mcitedefaultendpunct}{\mcitedefaultseppunct}\relax
\EndOfBibitem
\bibitem[McDermott \emph{et~al.}(2013)McDermott, Amelang, Lopatina, Reichhardt,
  and Reichhardt]{McDermott13}
D.~McDermott, J.~Amelang, L.~M. Lopatina, C.~J.~O. Reichhardt and
  C.~Reichhardt, \emph{Soft Matter}, 2013, \textbf{9}, 4607--4613\relax
\mciteBstWouldAddEndPuncttrue
\mciteSetBstMidEndSepPunct{\mcitedefaultmidpunct}
{\mcitedefaultendpunct}{\mcitedefaultseppunct}\relax
\EndOfBibitem
\bibitem[Feilhauer \emph{et~al.}()Feilhauer, Saha, Tobik, Zelent, Heyderman,
  and Mruczkiewicz]{Feilhauer19}
J.~Feilhauer, S.~Saha, J.~Tobik, M.~Zelent, L.~J. Heyderman and
  M.~Mruczkiewicz, {arXiv:1910.07388}\relax
\mciteBstWouldAddEndPuncttrue
\mciteSetBstMidEndSepPunct{\mcitedefaultmidpunct}
{\mcitedefaultendpunct}{\mcitedefaultseppunct}\relax
\EndOfBibitem
\end{mcitethebibliography}
  \bibliographystyle{rsc}
  }
\end{document}